\documentclass[%
 reprint,
 amsmath,amssymb,
 aps,
]{revtex4-2}

\usepackage{graphicx}
\usepackage{dcolumn}
\usepackage{bm}
\usepackage{hyperref}

\usepackage{comment}
\usepackage[dvipsnames]{xcolor}
\usepackage{orcidlink}

\hypersetup{
  colorlinks=true,
  linkcolor=red,
  citecolor=blue,
  urlcolor=blue,
 hypertexnames=false 
}

\newcommand{\GeV}{{\, \rm GeV}}

\begin{document}

\hfill TTP25-019,  P3H-25-041

\title{$B\rightarrow K + \text{axion-like particles}$: effective versus UV-complete models\\ and enhanced two-loop contributions}

\author{Xiyuan Gao\,\orcidlink{0000-0002-1361-4736}, and Ulrich Nierste\,\orcidlink{0000-0002-7442-4776}}
\email{xiyuan.gao@kit.edu,\\ ulrich.nierste@kit.edu}
\affiliation{\normalsize \it 
 Institut f\"ur Theoretische Teilchenphysik (TTP),
  Karlsruher Institut f\"ur Technologie (KIT), 76131 Karlsruhe, Germany.}

\begin{abstract}
An axion-like particle $a$ (ALP) can explain the excess of $B\rightarrow K + \text{invisible}$ events at Belle-II. However, many analyses of ALP scenarios are over-simplified. We revisit the $B\rightarrow K a$ transition rate in a popular minimal and UV complete model with two Higgs doublets (2HDM) and a complex singlet (DFSZ model). To this end we compare our results with previous studies which derived the $\overline{b}sa$ vertex from the $\overline{b}sA$ vertex, where $A$ is the heavy pseudo-scalar of the 2HDM, in terms of an $a-A$ mixing angle. 
We find this approach to work only at the leading one-loop order, while it fails at the two-loop level. Furthermore, while an approximate $Z_2$ symmetry suppresses the 
leading-order amplitude by a factor of 
$1/\tan\beta$, which is the ratio of the two vacuum expectation values of the Higgs doublets, we find the two-loop contribution unsuppressed and phenomenologically relevant for $\tan\beta \gtrsim 5$. We determine the allowed parameter space and underline the importance of better searches for  
$\Upsilon\rightarrow \gamma+$invisible and for a possible excess in $B\rightarrow K\mu^+\mu^-$.
We further study the low-energy axion effective theory which leads to a divergent and basis-dependent amplitude. As a conceptual result, we clarify the ambiguities and identify which low-energy framework is consistent with the DFSZ model.
\end{abstract}

\maketitle


\noindent

\textit{Introduction.}~~~~Recently Belle II has reported an excess in $B^+ \to K^+ \nu \overline{\nu}$ over the Standard Model (SM) prediction with a significance of $2.8 \sigma$~\cite{Belle-II:2023esi}. Since this excess is rather localized in the visible kaon energy,  a fit under the assumption of a two-body decay $B^+ \to K^+ a$ with invisible $a$ also gives an excellent fit to the data and has been performed in Ref.~\cite{Altmannshofer:2023hkn}. Using Belle II data the authors obtain a significance of $3.6 \sigma$ for $m_a \approx 2 \GeV$ and ${\rm BR} (B^+ \to K^+ a) = (8.8 \pm 2.5) \times 10^{-6}$. In a global analysis of Belle II and BaBar data the significance is reduced to about $2.4 \sigma$ with ${\rm BR} (B^+ \to K^+ a) = (5.1 \pm 2.1) \times 10^{-6}$.

However, the lightness of $a$ remains puzzling. A common interpretation is that $a$ is the pesudo-Goldstone boson of a spontaneously broken global symmetry. When the symmetry is the Peccei-Quinn $U(1)_{\text{{PQ}}}$~\cite{PQ1,PQ2} (with a soft breaking term), $a$ is called an axion-like particle (ALP). Since a renormalizable theory with only $a$ and SM particles is inconsistent, many related discussions are thus based on effective theories, see for example~Refs.~\cite{Batell:2009jf, Izaguirre:2016dfi, Aloni:2018vki,Chakraborty:2021wda, Gavela:2019wzg, Bauer:2021mvw, Calibbi:2025rpx, MartinCamalich:2025srw}. Although this is sufficient to explain the data, it does not allow an unambiguous connection between $b\rightarrow s a$ and other processes, such as $a$ decaying to SM particles. This is because, in the most general axion effective theory (axion-EFT)~\cite{Georgi:1986df}, the $b-s-a$ coupling strength is a free parameter. Correlations to the other interactions are, thus, not predicted~\footnote{
The interpretation with renormalizable group (RG) flow~\cite{Choi:2017gpf, Bauer:2020jbp} cannot relax this problem, due to the arbitrary boundary condition at ultraviolet (UV).}. Therefore, a complete renormalizable model is needed.

The minimal benchmark, DFSZ model~\cite{DFSZ1,DFSZ2}, was discussed in Ref.~\cite{Freytsis:2009ct}, where the authors assumed $b\rightarrow s a $ is induced by the mixing angle $\theta$ in $a-A$ mass term:
\begin{equation}
    \label{mixing}
    \mathcal{A}(b\rightarrow s a)~=~-\sin{\theta}\times\mathcal{A}(b\rightarrow s A).
\end{equation}
Here $A$ is a heavy CP-odd scalar of Type-II 2HDM~\cite{Glashow:1976nt,Branco:2011iw}, where $\mathcal{A}(b\rightarrow s A)$ is a finite one-loop effect~\cite{Wise:1980ux, Hall:1981bc, Frere:1981cc}. But we notice Eq.~(\ref{mixing}) requires clarification: when $A$ is heavier than the $b$ quark, the $b\rightarrow s A$ amplitude is off-shell and therefore unphysical and gauge-dependent. In fact, gauge-dependence challenges the entire mixing picture due to the subtlety of the $aG^+H^-$ vertex, as in the Higgs portal model discussed in Ref.~\cite{Kachanovich:2020yhi}. Another problem is that $\mathcal{A}(b\rightarrow s a)$ is suppressed when $\tan\beta$, the ratio of the two vacuum expectation values (VEVs) of 2HDM, is sizable. We find that $\mathcal{A}(b\rightarrow s a)$ can avoid the $\tan\beta$ suppression through higher-loop effects. So clearly, one needs to revisit $\mathcal{A}(b\rightarrow s a)$ including all physical and auxiliary states, and compare the results from the $a-A$ mixing picture provided in Ref.~\cite{Freytsis:2009ct}.

After clarifying the DFSZ model prediction, the next question is whether $B\rightarrow K a$ can be described without specifying UV physics. It's known that the $W^{\pm},G^{\pm}$ bosons contribute to $b\rightarrow s a$ with a divergent amplitude. So for a theory without tree-level $a$ flavor violating couplings, the effect of UV physics, whatever it is, must cancel the divergence and yield a model independent leading-log term $\log(\Lambda_{\text{UV}}/m_W)$. Interestingly, Ref.\cite{Dolan:2014ska} noted that the leading-log result differs by a factor of 4 in two reasonable operator bases: i) $a\overline{q}\gamma_5 q$ and ii) $\partial_{\mu}\overline{q}\gamma^{\mu}\gamma_5 q$. We will show that only ii) gives a leading-log result consistent with DFSZ model, and explore the reason in more detail. It reflects the fact that the decoupling of heavy physics requires gauge invariance in the light sector, as discussed in~\cite{Senjanovic:1979yq} nearly 50 years ago. This behavior is subtle but not unique, and some very similar results were found for $\mu\rightarrow e\gamma$ decays~\cite{Chang:1993kw, Davidson:2016utf, Altmannshofer:2020shb}.

\textit{$b\rightarrow s a$ in the DFSZ Model}~~~~We start with the DFSZ model, a minimal renormalizable theory for invisible ALPs. The scalar fields and their related interactions read~\cite{DFSZ1,DFSZ2,DiLuzio:2020wdo}:
\begin{equation}
    \label{Fields}
\begin{aligned}
    \Phi_{\alpha}~&=~\left(
    \begin{array}{c}
        \phi_{\alpha}^-   \\
          v_{\alpha}+(\rho_{\alpha}+i \eta_{\alpha})/\sqrt{2}
    \end{array}\right), \quad  \alpha~=~u,d, \\
\Phi_s~&=~f+\frac{r_0+i a_0}{\sqrt{2}}.
\end{aligned}
\end{equation}
\begin{equation}
\begin{aligned}
\label{DFSZpotential}
    V_{\Phi}~=&~ \Tilde{V}_{\text{moduli}}(|\Phi_u|,|\Phi_d|,|\Phi_u \Phi_d|,|\Phi_s|)+\lambda \Phi_s^{2} \Phi_u^{~} \Phi_d^{\dagger} +\text{h.c.},\\
    \mathcal{L}_{Y}~=&~ Y_u \overline{Q_L}u_R \Phi_u +Y_d \overline{Q_L}d_R \widetilde{\Phi}_d+Y_e \overline{L_L}e_R \widetilde{\Phi}_u+\text{h.c.}
\end{aligned}
\end{equation}
Here, $\Tilde{V}_{\text{moduli}}$ contains all gauge invariant combination of $\Phi_u,\Phi_d,\Phi_s$ without phase dependence, and $\widetilde{\Phi}_d=i\sigma_2 \Phi_d^{\dagger}$. $Y_u, Y_d, Y_e$ are $3\times3$ matrices, and the generation indices are implicit. 

To get insights, let us first focus on the limit $\lambda=0$, which is the PQWW model~\cite{PQ1,PQ2,WW1,WW2} plus a non-interacting complex singlet scalar. In this case, the interaction of Eq.~(\ref{DFSZpotential}) admits three $U(1)$ symmetries, corresponding to independent phase rotations of $\Phi_u, \Phi_d, \Phi_s$. All $U(1)$ symmetries are spontaneously broken when the scalar fields take their VEVs, giving three massless Goldstone modes. One of them is gauged by the hypercharge $U(1)_Y$, and the other two are PQWW visible axion $A_{0}$ and the massless radial mode of $a_0$ from $\Phi_s$. The PQWW $b\rightarrow s A_0$ amplitude was analyzed by~\cite{Wise:1980ux, Hall:1981bc, Frere:1981cc}, and $b\rightarrow s a_0$ is clearly zero due to the $a_0\rightarrow -a_0 $ (or $\Phi_s\rightarrow\Phi_s^{*})$ symmetry. Then, by introducing a tiny $a_0-A_0$ mass matrix that only softly breaks the $U(1)$ symmetries but generates a physical mixing angle $\theta$, the amplitude for the physical state $a$ reads: 
\begin{equation}
    \label{LigetiResult}
    \mathcal{A}(b\rightarrow s a)_{\text{DFSZ}}~=~-\sin{\theta}\times \mathcal{A}(b\rightarrow s A_0)_{\text{PQWW}}.
\end{equation}
Eq.~(\ref{LigetiResult}) clarifies the meaning of Eq.~(\ref{mixing}) analyzed in Ref.~\cite{Freytsis:2009ct} in the limit $\lambda=0$. 

\begin{figure}
    \centering
    \includegraphics[width=0.43\textwidth]{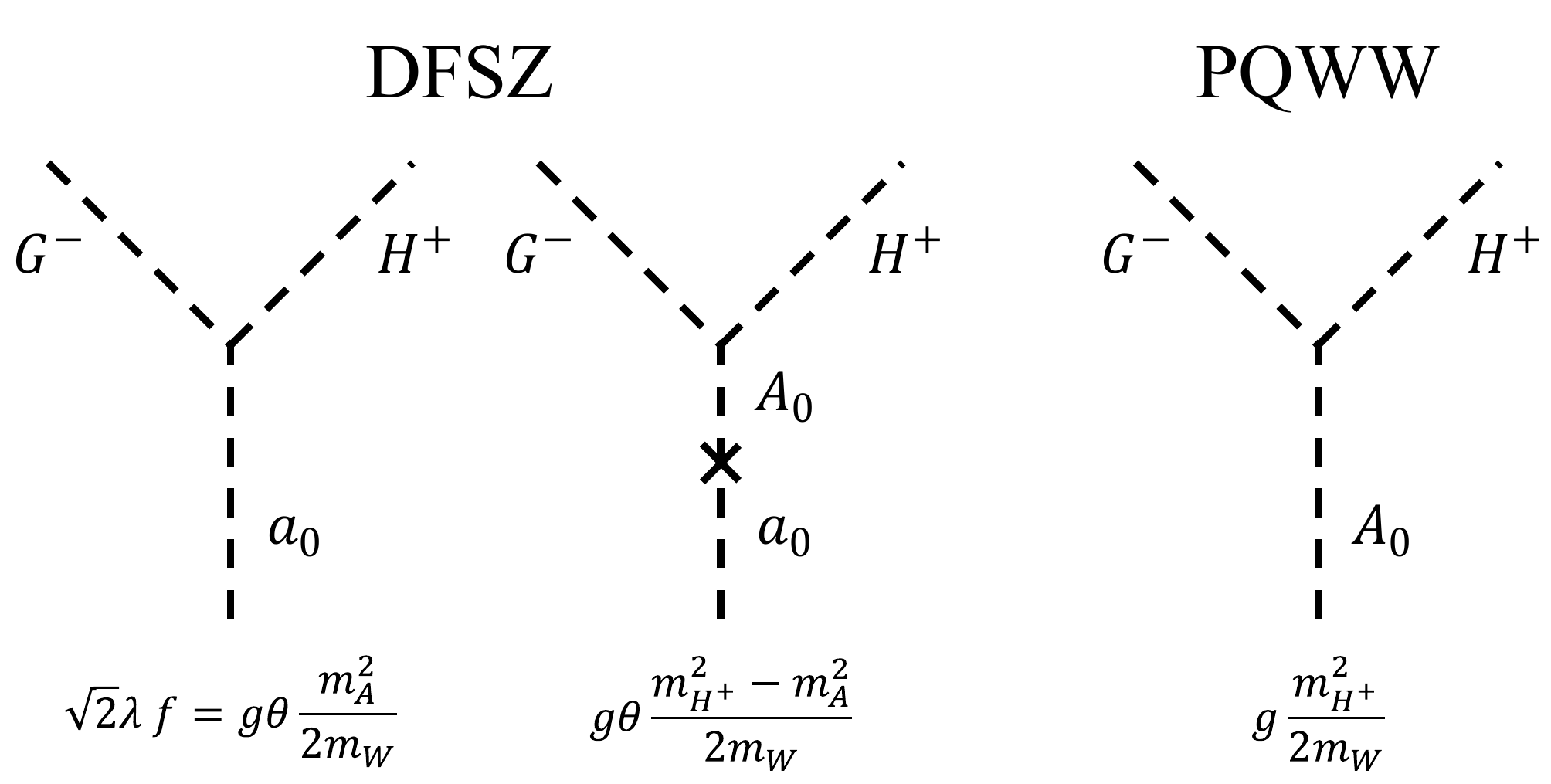}
    \caption{Feynrules for $G^-H^+-$ALP coupling in DFSZ (left) and PQWW (right) models.}
    \label{feynrules1}
\end{figure}

What if $\lambda=0$? It breaks global $U(1)_{\text{PQWW}} \times U(1)_{s}$ to $U(1)_{\text{PQ}}$. One of the $a_0-A_0$ mixing state, $A$, acquires large mass, and the $b\rightarrow s A_0$ amplitude could be off shell. In addition, some Feynman rules change: The $G^-H^+A_0$ vertex rule is modified by a $m_{A}^2$ term, and a new $G^-H^+a_0$ vertex appears, as shown in Fig.~\ref{feynrules1}. Therefore, we conclude: 
\begin{equation}
\begin{aligned}
    \mathcal{A}(b\rightarrow s a)_{\text{DFSZ}}~&\neq~-\sin{\theta}\times\mathcal{A}(b\rightarrow s A_0)_{\text{DFSZ}}, \\
    \mathcal{A}(b\rightarrow s A_0)_{\text{DFSZ}}~&\neq~\mathcal{A}(b\rightarrow s A_0)_{\text{PQWW}}.\\
\end{aligned}
\end{equation}
And strictly speaking, the off-shell amplitude $\mathcal{A}(b\rightarrow s A_0)_{\text{DFSZ}}$ is not physical since its gauge dependent. However, we find the Feynman rule of the effective $G^-H^+a_0$ vertex remains the same as the $G^-H^+A_0$ one of PQWW multiplied by $\theta$. As illustrated in Fig.~\ref{feynrules1}, the $m_A^2$ term cancels and Eq.~(\ref{LigetiResult}), the explicit expression of Ref.~\cite{Freytsis:2009ct}, is correct. We think the change is not accidental. After all, the $\lambda \Phi_s^{2} \Phi_u^{~} \Phi_d^{\dagger}$ term does not contain the physical state $a$, when parameterized exponentially. In the non-linear basis, the $G^-H^+a$ vertex becomes $G^-\overset{\leftrightarrow}{\partial^{\mu}}H^+ \partial_{\mu}a$, and is manifestly independent of $m_A$.

The $\lambda \Phi_s^2\Phi_u\Phi_d$ term implies that $a$ can directly interact with the scalar fields and the mixing picture breaks down unless $\lambda=0$. However, $\lambda$ does not appear in the explicit expression. This can be understood by analyzing the symmetry in the broken phase. In the limit $v_d\ll v_u\ll f$, all mixing angles are suppressed, aligning the mass and interaction basis for the scalar fields:
\begin{equation}
\label{notionZ2}
\begin{aligned}
    \Phi_u&\simeq H_u= 
    \left(\begin{array}{c}
        G^-   \\
          v+\frac{h+i G^0}{\sqrt{2}}
    \end{array}\right),~
    \Phi_d\simeq H_d=\left(\begin{array}{c}
        H^-   \\
          \frac{H+i A}{\sqrt{2}}
    \end{array}\right),\\
    \Phi_s&\simeq f+\frac{r+i a}{\sqrt{2}}. 
\end{aligned}
\end{equation}
If in addition $\lambda\ll1$, the following $Z_2$ symmetry emerges in $V_{\Phi}$ and $\mathcal{L}_{Y}$: 
\begin{equation}
    \label{Z2sym}
    d_R\rightarrow-d_R, \quad H_d\rightarrow-H_d, \quad  \text{others unchanged.}
\end{equation}
Let's define:
\begin{equation}
\label{alignmentLimit}
\begin{aligned}
    \theta\equiv\frac{2 v_d}{f}, \quad \frac{1}{\tan{\beta}}\equiv\frac{v_d}{v_u}, \quad  \lambda
\end{aligned}
\end{equation}
as $Z_2$ spurions, the effective Hamiltonian for $b\rightarrow s a$ is then fixed up to the loop factors $X_1,X_2$ and $X_3$:
\begin{equation}
\label{Hamilton}
\begin{aligned}
    \mathcal{H}_{\text{eff}}~=&~ \theta \frac{g^3 V_{ts}^*V_{tb}}{128\pi^2} \frac{m_t^2}{m_W^3}
    \overline{s}\gamma^{\mu}P_Lb~\partial_{\mu} a\\
    &\times
    \left(
    X_1\frac{1}{\tan\beta}+X_2\frac{1}{\tan^3\beta}+X_3\frac{\lambda}{16\pi^2}
    \right).
\end{aligned}
\end{equation}
$X_1$ and $X_2$ are given in Ref.~\cite{Freytsis:2009ct}, while $\lambda$ enters the expression only with an additional loop factor $1/16\pi^2$. This is because $\lambda$ carries Planck Units $\hbar$. As a coupling constant, $\lambda$ arises only at the next order in the perturbative expansion. This is why its contribution is absent in the one-loop expression of Ref.~\cite{Freytsis:2009ct}. We notice the one-loop result is suppressed when $\tan\beta$ is sizable, so the $\lambda$ contribution can be important.

Let's define $m_H\equiv m_{H^+}$ and take the limit that $m_H^2\simeq m_A^2=\lambda f^2\tan\beta \gg m_W^2$, the loop factors in Eq.~(\ref{Hamilton}) reads:
\begin{equation}
\begin{aligned}
\label{X1X2X3}
    X_1~=&~-\log{\frac{m_H^2}{m_{t}^2}}+\frac{3 m_W^4}{(m_t^2-m_W^2)^2}\log{\frac{m_t^2}{m_W^2}}\\
    &+\frac{3(m_t^2-2m_W^2)}{m_t^2-m_W^2},\\
    X_2~=&~0,\\
    X_3~=&~\log{\frac{m_{H}^2}{m_t^2}}+\frac{6m_W^2}{m_t^2-m_W^2}\log{\frac{m_t^2}{m_W^2}}+\frac{1}{2}.\\
\end{aligned}
\end{equation}
Our results for $X_1$ and $X_2$ agree with Ref.~\cite{Freytsis:2009ct} in the large $m_H$ limit. $X_3$ is new and  could in principle contain transcendental-weight-two functions, but only the logarithm functions appear after expanding in $1/m_H$. 

\begin{figure}[t!]
    \centering
    \includegraphics[width=0.48\textwidth]{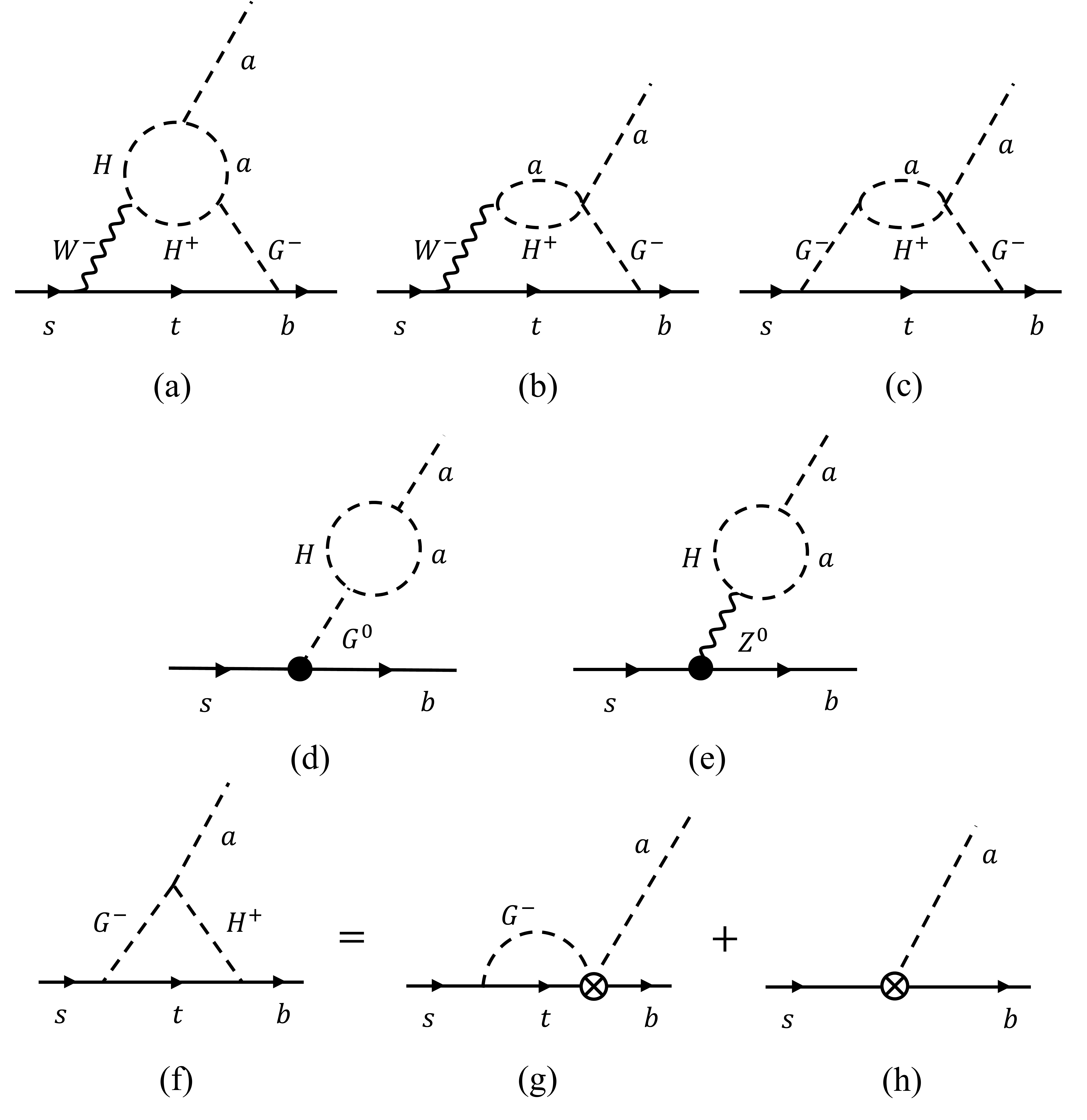}
    \caption{Some Feynman diagrams related to the discussion, where the black dots in (d) and (e) indicate one-loop vertices and the crossed dots in (g) and (h) indicate the effective vertices after integrating out $H^+$.
    }
    \label{feynrules2}
\end{figure}

To calculate $X_3$, we work in the general $R_{\xi}$ gauge, to check our results by showing $\xi$ independence. We use the packages FeynRules~\cite{Christensen:2008py,Alloul:2013bka} and FeynArts~\cite{Hahn:2000kx} to generate the new interacting vertexes and new diagrams. In total thousands of diagrams contribute to $b\rightarrow s a$ at two-loop level, but only very few contribute to $X_3$, as shown in Fig.~\ref{feynrules2}. For simplicity, the diagrams from exchanging $b\leftrightarrow s$, and the ones replacing the internal $a,H$ with $r,A$~\footnote{$m_r\ll m_A$ in the large $\tan\beta$ limit.}, are not shown explicitly. The others come with additional factor of $\tan\beta$ and $\theta$ are not relevant.

To evaluate the diagrams, we Taylor expand the Feynman amplitudes in the small external momentum $p_i^{\mu}\sim m_b$~\cite{Fleischer:1994ef}: 
\begin{equation}
    \label{Tayler}
    \mathcal{A}~=~\mathcal{A}|_{p_i^{\mu}=0}+p_i^{\mu}\left.\frac{\partial \mathcal{A}}{\partial p_i^{\mu}}\right|_{p_i^{\mu}=0}+\mathcal{O}(p_i^2).
\end{equation}
In the large $\tan\beta$ limit, we find the $\mathcal{A}|_{p_i^{\mu}=0}$ do not contribute to $X_3$. Its effect is canceled after renormalizing the quark-mixing matrix, which arise from the same diagrams by replacing the external $a$ with the vacuum tadpole $f$. So only $\left.\frac{\partial \mathcal{A}}{\partial p_i^{\mu}}\right|_{p_i^{\mu}=0}$ is relevant, and we evaluate it using FeynCalc~\cite{Shtabovenko:2016sxi,Shtabovenko:2020gxv, Mertig:1990an}, FIRE~\cite{Smirnov:2019qkx}, and FeynHelpers~\cite{Shtabovenko:2016whf}. The new functions for multiloop tensor reduction and topological identification in Feyncalc 10~\cite{Shtabovenko:2023idz} are applied. The master integrals are reduced to the vacuum bubble ones, whose analytical expressions are given in~\cite{Davydychev:1992mt,Nierste:1995fr}.

The cancellation of the gauge parameter $\xi_W$ is similar to the case of $b\rightarrow s \mu^+\mu^-$ in SM~\cite{Inami:1980fz}. Intuitively, the $\xi_W$ dependent contribution from the box-like diagrams (a-c) cancel those from the penguin-like diagrams (d-e). Thus, the loop-induced kinetic-mixing term $\partial_{\mu}G^0\partial^{\mu}a$ 
plays an important role. To perform $a-G^0$ wave-function renormalization, we apply the standard method of subtracting the Goldstone boson self-energies at zero momentum~\cite{Santos:1996vt, Grinstein:2015rtl}. This eliminates the tadpole contributions and simplifies the diagrams.

\textit{$b\rightarrow s a$ without specifying UV physics}~Is it possible to find the properties of the $b\rightarrow s a$ amplitude of the DFSZ model from an effective theory, without specifying any UV physics? Naively, one can start with the renormalizable Lagrangian of Eq.~(\ref{DFSZpotential}) but drop all heavy particles: 
\begin{equation}
    \label{lightsubtheory}
    \mathcal{L}~=~ \mathcal{L}_{\text{SM}}~+~ i a \sum_{q=t,b} c_{q}~\overline{q} \gamma_5 q. 
\end{equation}
This looks quite reasonable, since the light ALP $a$ only changes the IR structure of the SM. The low-energy effective field theory (LEFT)~\cite{Jenkins:2017jig,Jenkins:2017dyc} respects the QED$\times$QCD symmetry, under which the quarks and leptons are vector-like. Therefore, the $a$-fermion couplings can be renormalizable. With $c_{q}$ matched from the DFSZ model, one can calculate the $b\rightarrow s a$ amplitude, with UV divergence. Applying the RG equations, one finds the leading-log term appears: 
\begin{equation}
\label{leadinglogtermR}
    \mathcal{A}(b\rightarrow s a)~\sim~ \log {(\Lambda_{\text{UV}}^2/m_t^2)}.
\end{equation}
In the DFSZ model $\Lambda_{UV}$ is equal to $m_H$~\footnote{As pointed out in \cite{Alonso-Alvarez:2021ett}, $\Lambda_{UV}$ is not necessarily equal to the ALP decay constant $f$. The UV cut-off scale is the one above which the heavy particles can no longer be integrated out.}, however, we find the coefficient of Eq.~(\ref{leadinglogtermR}) in disagreement with the $X_1$ term of Eq.~(\ref{X1X2X3}). Thus, something is wrong. 

We carefully checked the DFSZ calculation and found the missing term comes from Fig.~\ref{feynrules2}(f). The low-energy theory can not capture its contribution because the Feynman rules of the $G^-H^+a$ vertex is proportional to $m_H^2$. Fig.~\ref{feynrules2}(f) is not suppressed by $1/m_H^2$ although it contains a heavy particle. Although $H^+$ is much heavier than $W^+$, they are equally important in $b\rightarrow s a$. This challenges the naive understanding about IR-UV mixing/decoupling.

\begin{figure*}[t!]
  \centering
  \includegraphics[width=0.95\textwidth]{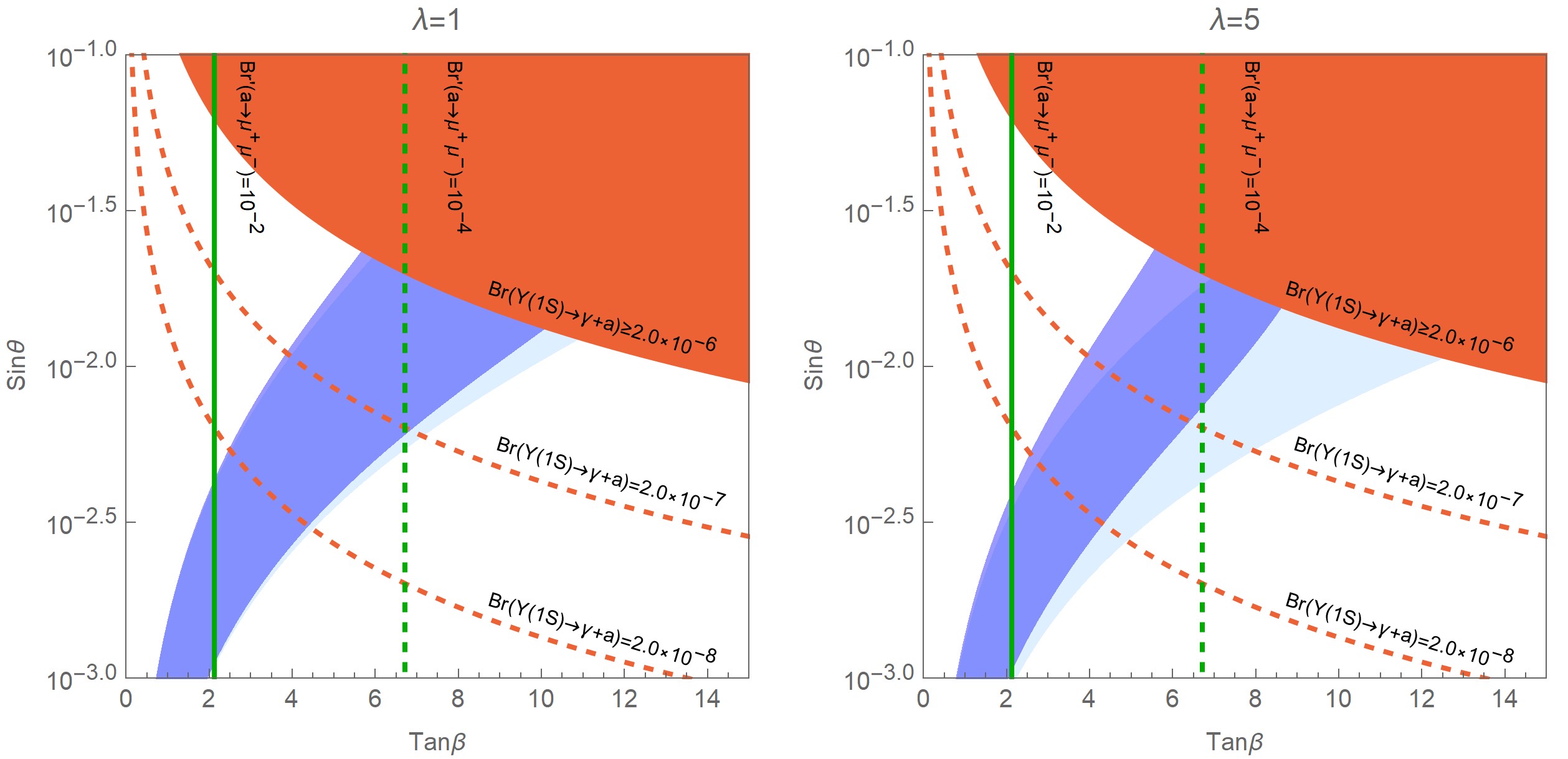}
    \caption{Parameter space explaining the Belle II excess in the DFSZ model, for $\lambda=1$ (left) and $\lambda=5$ (right). The dark (light) blue region gives Br$(B\rightarrow K a)=(1\sim 9)\times 10^{-6}$, with complete (one loop only) calculation. The red region is excluded by the search for $\Upsilon(nS)\rightarrow \gamma a$. The green contours indicate the visible branching ratio $\text{Br}'(a\rightarrow \mu^+\mu^-)=\text{Br}(a\rightarrow \mu^+\mu^-)/\text{Br}(a\rightarrow \text{visible})$.}
   \label{DFSZII}
\end{figure*}

In our opinion, the reason is that without $H^+$, the renormalizable theory of Eq.~(\ref{lightsubtheory}) is not invariant under $SU(2)_L\times U(1)_Y$. The necessary condition for decoupling, that the light theory must be gauge invariant~\cite{Senjanovic:1979yq}, is not satisfied. Decoupling is hidden in the dimensionless coupling:
\begin{equation}
\label{Nondecoupling}
    c_{q}~\sim\theta~\sim~\frac{1}{f} ~\sim~ \frac{1}{m_H}. 
\end{equation}
Here, $m_H$ can not be arbitrarily heavy given finite $c_{q}$. This behavior is somehow uncommon, but not unique. For instance, the $\mu\rightarrow e  \gamma$ decay amplitude in a 2HDM is not directly suppressed by the heavy mass either~\cite{Chang:1993kw}, but by the misalignment parameter $c_{\alpha\beta}\sim 1/m_H^2$~\cite{Gunion:2002zf}. With finite $c_{q}$ or $c_{\alpha\beta}$, one can not recover $SU(2)_L\times U(1)_Y$.

To reveal the decoupling picture, the $SU(2)_L\times U(1)_Y$ gauge symmetry must be respected by the low energy theory. It is chiral, unlike QED$\times$QCD. The $SU(2)$ doublet Higgs must join the low-energy-theory of Eq.~(\ref{lightsubtheory}) so renormalizability cannot hold anymore. The gauge invariant Lagrangian reads:
\begin{equation}
\begin{aligned}
    \label{EFTS2}
    \mathcal{L}
     ~=&~\mathcal{L}_{\text{SM}}+i \frac{a}{v}\left( c_b \overline{Q}_L  b_R \widetilde{H}_u+c_t ~\overline{Q}_Lt_RH_u+\text{h.c.}  \right) \\
    =&~\mathcal{L}_{\text{SM}}+i a \sum_{q=t,b} c_{q}\overline{q} \gamma_5 q
    + i \frac{a}{v}\left[ c_b V_{tb}^{}  \overline{t}_L  b_R G^+\right.\\&\left.+c_t \left(V_{tb}^{*}\overline{b}_L t_R G^-+ V_{ts}^{*}\overline{s}_L t_R G^-\right)+\text{h.c.}  \right]+...
\end{aligned}
\end{equation}
The key difference is the appearance of non-renormalizable operators with unphysical Goldstone Mode $G^+$. They have a clear UV origin, as illustrated in Fig~\ref{feynrules2}(g). Splitting the propagator of $H^-$ into two pieces~\cite{Bilenky:1993bt},
\begin{equation}
\label{splitting}
    \frac{1}{k^2-m_H^2}~=~-\frac{1}{m_H^2}+\frac{1}{m_H^2}\frac{k^2}{k^2-m_H^2},
\end{equation}
the $-1/m_H^2$ term leads to the non-renormalizable operator, while the $m_H$ dependence is canceled since the $G^-H^+a$ vertex is proportional $m_H^2$. This effective operator, as shown Fig.~\ref{feynrules2}(g), leads to a divergent amplitude, and we checked that it exactly reproduces the leading-log term missing in Eq.~(\ref{leadinglogtermR}). As previously discussed, the light theory of the 2HDM is also not gauge invariant. Very similar operators with Goldstone bosons contribute to $\mu\rightarrow e\gamma$ with a leading-log term. We refer the reader to Ref.~\cite{Altmannshofer:2020shb}, for details about this closely related example.

If one picks the unitary gauge, Eq.~(\ref{EFTS2}) and Eq.~(\ref{lightsubtheory}) are the same, since the gauge fixing condition sets $G^+(x^{\mu})\equiv 0$. We have checked that the missing leading-log term of Eq.~(\ref{leadinglogtermR}) now originates from the the longitudinal part of $W$ propagator $k_{\mu}k_{\nu}/m_W^2$. Decoupling works, because the gauge symmetry is strictly speaking still preserved, just hidden by gauge fixing. And again, the cost is loosing renormalizability, known as a consequence of the unitary (non-renormalizable) gauge.

By applying the equations of motions, Eq.~(\ref{EFTS2}) becomes the general axion EFT where the $U(1)_{\text{PQ}}$ symmetry is manifest~\cite{Georgi:1986df, Choi:2017gpf, Bauer:2020jbp}: 
\begin{equation}
\begin{aligned}
    \label{EFTS}
    \mathcal{L}~=&~\mathcal{L}_{\text{SM}}+\sum_{\psi_L=Q_L,t_R^c, b_R^c}\frac{c_{\psi}}{f}~\overline{\psi}_L\gamma^{\mu} \psi_L ~\partial_{\mu}a+...
\end{aligned}
\end{equation}
Anomalous terms such as  $a\tilde{W}_{\mu\nu}W^{\mu\nu}$~\cite{Bauer:2020jbp} are higher order for flavor violating processes~\cite{Izaguirre:2016dfi}, so we don't show them explicitly here. Clearly, this derivative basis produces the same $b\rightarrow s a$ amplitude as the one of Eq.~(\ref{EFTS2}). However, the Yukawa basis of Eq.~(\ref{lightsubtheory}) gives a different result. The authors of Ref.~\cite{Dolan:2014ska} have commented on this discrepancy in a footnote and correctly connected it to the dimension-5 operators. Here, we emphasize that Eq.~(\ref{lightsubtheory}) is inconsistent without gauge fixing. 

Before finishing the bottom-up discussions, we want to emphasize that $\log {(\Lambda_{\text{UV}}^2/m_t^2)}$ is large and the terms without this  leading log are not available without specifying UV physics. UV physics is hidden in the counter term of Fig.~\ref{feynrules2}(h) (from the third term of Eq.~(\ref{splitting})). The general ALP effective theory allows tree level flavour violating couplings. Strictly speaking, $b\rightarrow s a$ itself is a \textit{definition} of the renormalization scheme, about how Fig.~\ref{feynrules2}(h) cancels the divergence of Fig.~\ref{feynrules2}(g), not a \textit{prediction} of the EFT.

\textit{Phenomenology}~The $b\rightarrow s a$ decay amplitude alone is not sufficient to explain the Belle II excess in a self-consistent way. The invisible signal requires $a$ to escape detection. However, the DFSZ model implies that $a$ is short-lived and decays inside the detector.
Explaining the Belle-II excess needs invisible $a$ decay channels, which is beyond the model prediction. The DFSZ model has to be extended with a dark sector, for example, a light sterile particle $\chi$ with $a\overline{\chi}\chi$ coupling only. Here, we do not try to build dark matter models, but assume $\text{Br}^{\text{inv}}\equiv \text{Br}(a\rightarrow \text{invisible})\approx 1$ for simplicity. If $\text{Br}^{\text{inv}}\ll1$, the required value for the mixing angle $\theta$ must be enhanced by a factor of $1/\sqrt{\text{Br}^{\text{inv}}}$.

We also checked the consistency with other search limits.
The various visible decay rates of a general axion-EFT are shown in Ref.~\cite{Bauer:2017ris}. 
Considering the detecting limits, the only relevant visible channel is the charged lepton one, $a\rightarrow\mu^+\mu^-$. It contributes an excess in $B\rightarrow K \mu^+\mu^-$ at low $q^2\sim$ 2 GeV. If the total visible decay rate is non-negligible, escaping the current limits requires sizable $\tan\beta$ so that $H_d$ weakly couples to leptons. On the other hand, very large $\tan\beta$ is excluded by $\Upsilon(1S)\rightarrow \gamma+\text{invisible}$ searches. The current limit disfavors the $\tan\beta\gtrsim10$ region.




We illustrate the three phenomenologically relevant processes in Fig.~\ref{DFSZII}, in the benchmark scenario $\lambda=1$ and $\lambda=5$. The dark blue regions represent the values of $\sin\theta$ and $\tan\beta$ that can explain the Belle-II excess. The light blue regions are based on one-loop calculations alone and mostly overlap with the dark ones. However, they differ when both $\tan\beta$ and $\lambda$ are sizable enough. In this case, the two-loop calculation we newly computed in this work becomes important. Notably, the two-loop and one-loop amplitudes have opposite sign and partly cancel, so favoring a larger value of $\sin\theta$. 
Assuming a sizable visible branching ratio, the viable $\sin\theta-\tan\beta$ parameter space to explain the Belle-II excess becomes fully constrained. Consequently, either a $B\rightarrow K\mu^+\mu^-$ excess or  $\Upsilon\rightarrow \gamma+$invisible should be observed in future experiments. Detection can only be avoided if the total visible decay rate is negligible. 

\textit{Conclusion and Discussion}~~~~We revisited the $B\rightarrow K a$ transition rate in the DFSZ model, which is a minimal UV-complete benchmark for an ALP $a$. Studying approximate symmetries suppressing the known one-loop amplitudes, we determine new unsuppressed two-loop contributions. When $\tan\beta$ is sizable, our result becomes essential. In addition, while it is possible to capture the key features of DFSZ model with a bottom-up approach, the choice of the low energy theory is subtle. Only the gauge invariant EFT yields the correct leading-log term, with the cost of loosing renormalizability. 


From a practical side of view, we agree that the operator $a\overline{s}\gamma_5 b$ alone is sufficient to explain the Belle-II excess. 
However, the DFSZ model should be taken more seriously as a minimal benchmark for light new particles with minimal flavor violation (MFV)~\cite{Chivukula:1987py, DAmbrosio:2002vsn}. Here, the rare $B\rightarrow K a$ decay rate is suppressed by a loop factor $1/(16\pi^2)$, small flavor mixing angles, and a possible hierarchy between two VEVs. So the new physics for UV completion needs not be super-heavy, but could be in the TeV range. In other words, some other beyond-SM processes should not be far away from detection. Therefore, it is reasonable to expect detecting $\Upsilon\rightarrow \gamma +a$ and $B\rightarrow K a\rightarrow K \mu^+\mu^-$ signals in future experiments, which will support the model.

\begin{acknowledgments}

We are grateful to Robert Ziegler for useful discussions and comments. This research was supported by the Deutsche Forschungsgemeinschaft (DFG, German Research Foundation) under grant 396021762 - TRR 257 and by the BMBF Grant 05H21VKKBA, \textit{Theoretische Studien für Belle II und LHCb.} X.G. also acknowledges the support by the Doctoral School ``Karlsruhe School of Elementary and Astroparticle Physics: Science and Technology.''

\end{acknowledgments}

\nocite{*}

\bibliography{apssamp}

\newpage

\appendix

\end{document}